\documentclass[final,1p,times]{elsarticle}
\usepackage{amssymb}
\usepackage{amsfonts}
\usepackage{amsmath}
\usepackage{geometry}
\usepackage{bm}% bold math
\newtheorem{thm}{Theorem}

\newdefinition{rmk}{Remark}
\newproof{pf}{Proof}
\newproof{pot}{Proof of Theorem}
\begin{document}

\begin{frontmatter}
\title{The power series expansion of Mathieu function\\ and its integral formalism }% Force line breaks with \\
%\thanks{Footnote to title of article.}

\author{Yoon Seok Choun\corref{cor1}}
\ead{Yoon.Choun@baruh.cuny.edu; ychoun@gc.cuny.edu; ychoun@gmail.com}
\cortext[cor1]{Correspondence to: Baruch College, The City University of New York, Natural Science Department, A506, 17 Lexington Avenue, New York, NY 10010} 
\address{Baruch College, The City University of New York, Natural Science Department, A506, 17 Lexington Avenue, New York, NY 10010}
\begin{abstract}

Mathieu ordinary differential equation is of Fuchsian types with the two regular and one irregular singularities. In contrast, Heun equation of Fuchsian types has the four regular singularities. Heun equation has the four kind of confluent forms: (1) Confluent Heun (two regular and one irregular singularities), (2) Doubly confluent Heun (two irregular singularities), (3) Biconfluent Heun (one regular and one irregular singularities), (4) Triconfluent Heun equations (one irregular singularity).  For DLFM version \cite{NIST}, Mathieu equation in algebraic forms is also derived from the Confluent Heun equation by changing all coefficients $\delta =\gamma =\frac{1}{2}$, $\epsilon =0$, $\alpha =q$ and $q=\frac{\lambda +2q}{4}$.

In this paper I apply three term recurrence formula \cite{chou2012b} to the power series expansion in closed forms of Mathieu equation for infinite series and its integral forms including all higher terms of $A_n$'s. 
One interesting observation resulting from the calculations is the fact that a modified Bessel function recurs in each of sub-integral forms: the first sub-integral form contains zero term of $A_n's$, the second one contains one term of $A_n$'s, the third one contains two terms of $A_n$'s, etc. Section 5 contains two additional examples of Mathieu function. 

This paper is 5th out of 10 in series ``Special functions and three term recurrence formula (3TRF)''. See section 6 for all the papers in the series. Previous paper in series deals with asymptotic behavior of Heun function and its integral formalism \cite{Chou2012d}. The next paper in the series describes the power series expansion in closed forms of Lame equation in the algebraic form and its integral forms \cite{Chou2012f}.
\end{abstract}

\begin{keyword}
Mathieu equation, Integral form, Three term recurrence formula, Modified Bessel function

%\PACS{02.10.De, 02.30.Hq, 02.30.Ik, 02.30.Jr, 02.30.Gp, 02.30.Mv, 03.65.Ge, 03.65.-w, 03.65.Fd, 04.62.+v, 04.60.-m, 04.65.+e}
\MSC{33E10, 34B30}
\end{keyword}
                                      
\end{frontmatter}  
\section{Introduction}

One example of three therm recursion relations is the Mathieu equation, introduced by \'Emile L\'eonard Mathieu (1868)\cite{Math1868}, while investigating the vibrating elliptical drumhead. Mathieu equation, known for the elliptic cylinder equation, appears in diverse areas such as astronomy and physical problems involving Schr$\ddot{\mbox{o}}$dinger equation for a periodic potentials \cite{Conn1984}, the parametric Resonance in the reheating process of universe \cite{Zlat1998}, and wave equations in general relativity\cite{Birk2007}, etc. Mathieu function has been used in various areas in modern physics and mathematics. \cite{McLa1947,Guti2003,Daym1955,Troe1973,Alha1995,Shen1981,Bhat1988,Ragh1991,Aliev1999} 

Unfortunately, even though Mathieu equation has been observed in various areas mentioned above, there are no power series expansion in closed forms and its integral formalism in analytically.  
The Mathieu function has only been described in numerical approximations (Whittaker 1914\cite{Whit1914}, Frenkel and Portugal 2001\cite{Fren2001}). Sips 1949\cite{Sips1949}, Frenkel and Portugal 2001\cite{Fren2001} argued that it is not possible to represent analytically the Mathieu function in a simple and handy way.

I construct the power series expansion of Mathieu equation in closed forms and its integral representation analytically using three-term recurrence formula (3TRF)\cite{chou2012b} and the same method I used in analyzing Heun function.\cite{chou2012c,Chou2012d}

Mathieu equation is witten by
\begin{equation}
 \frac{d^2{y}}{d{z}^2} + \left( \lambda - 2q\;\mathrm{\cos}2z \right) y = 0 \label{eq:1}
\end{equation}
where $\lambda $ and $q$ are parameters. This is an equation with periodic-function coefficient. Mathieu equation also can be described in algebraic forms putting $x=\mathrm{\cos}^2z$:
\begin{equation}
 4x (1-x ) \frac{d^2{y}}{d{x}^2} + 2( 1-2x) \frac{d{y}}{d{x}} + ( \lambda + 2 q - 4 q x ) y = 0\label{eq:2}
\end{equation}
This equation has two regular singularities: $x=0$ and $x=1$; the other singularity $x=\infty $ is irregular. Assume that its solution is
\begin{equation}
y(x)= \sum_{n=0}^{\infty } c_n x^{n+\nu }\label{eq:3}
\end{equation}
where $\nu$ is an indicial root. Plug (\ref{eq:3}) into (\ref{eq:2}).
\begin{equation}
c_{n+1}=A_n \;c_n +B_n \;c_{n-1} \hspace{1cm};n\geq 1\label{eq:4}
\end{equation}
where,
\begin{subequations}
\begin{equation}
A_n = \frac{4(n+\nu )^2-(\lambda +2q)}{2(n+1+\nu )(2(n+\nu )+1)}\label{eq:5a}
\end{equation}
\begin{equation}
B_n = \frac{4q}{2(n+1+\nu )(2(n+\nu )+1)}\label{eq:5b}
\end{equation}
\begin{equation}
c_1= A_0 \;c_0 \label{eq:5c}
\end{equation}
\end{subequations}
We have two indicial roots which are $\nu = 0$ and $ \frac{1}{2} $. As we see (\ref{eq:5b}), there is no way to make $B_n$ term terminated at certain value of index $n$. Because the numerator of (\ref{eq:5b}) is just consist of constant $q$ parameter. So there are only two kind of power series expansions which are infinite series and polynomial which makes $A_n$ term terminated. In this paper I construct an analytic solution of Mathieu equation for infinite series. And in the future I will construct its analytic solution for polynomial which makes $A_n$ term terminated.\footnote{Its local solutions are available in chapter 7 of Ref.\cite{Choun2013}}

\section{Power series expansion for infinite series}
In Ref.\cite{chou2012b} the general expression of power series of $y(x)$ for infinite series is defined by
\begin{eqnarray}
y(x)  &=& \sum_{n=0}^{\infty } y_{n}(x)= y_0(x)+ y_1(x)+ y_2(x)+ y_3(x)+\cdots \nonumber\\
&=& c_0 \Bigg\{ \sum_{i_0=0}^{\infty } \left( \prod _{i_1=0}^{i_0-1}B_{2i_1+1} \right) x^{2i_0+\lambda } 
+ \sum_{i_0=0}^{\infty }\left\{ A_{2i_0} \prod _{i_1=0}^{i_0-1}B_{2i_1+1}  \sum_{i_2=i_0}^{\infty } \left( \prod _{i_3=i_0}^{i_2-1}B_{2i_3+2} \right)\right\} x^{2i_2+1+\lambda }  \nonumber\\
&& + \sum_{N=2}^{\infty } \Bigg\{ \sum_{i_0=0}^{\infty } \Bigg\{A_{2i_0}\prod _{i_1=0}^{i_0-1} B_{2i_1+1} 
 \prod _{k=1}^{N-1} \Bigg( \sum_{i_{2k}= i_{2(k-1)}}^{\infty } A_{2i_{2k}+k}\prod _{i_{2k+1}=i_{2(k-1)}}^{i_{2k}-1}B_{2i_{2k+1}+(k+1)}\Bigg)\nonumber\\
&& \times  \sum_{i_{2N} = i_{2(N-1)}}^{\infty } \Bigg( \prod _{i_{2N+1}=i_{2(N-1)}}^{i_{2N}-1} B_{2i_{2N+1}+(N+1)} \Bigg) \Bigg\} \Bigg\} x^{2i_{2N}+N+\lambda }\Bigg\} 
\label{eq:6}
\end{eqnarray}
Substitute (\ref{eq:5a})-(\ref{eq:5c}) into (\ref{eq:6}). In this article Pochhammer symbol $(x)_n$ is used to represent the rising factorial: $(x)_n = \frac{\Gamma (x+n)}{\Gamma (x)}$. The general expression of power series of Mathieu equation for infinite series is given by
\begin{eqnarray}
 y(x) &=& \sum_{n=0}^{\infty } y_n(x)= y_0(x)+y_1(x)+ y_2(x) + y_3(x)+\cdots\nonumber\\
&=& c_0 x^{\nu } \left\{\sum_{i_0=0}^{\infty} \frac{1}{(1+\frac{\nu}{2})_{i_0} (\frac{3}{4}+\frac{\nu}{2})_{i_0}} \eta^{i_0} +  \left\{\sum_{i_0=0}^{\infty } \frac{(i_0+\frac{\nu}{2})^2 - \frac{1}{4^2}(\lambda +2q)}{(i_0+\frac{1}{2}+\frac{\nu}{2})(i_0+\frac{1}{4}+\frac{\nu}{2})} \frac{1}{(1+\frac{\nu}{2})_{i_0} (\frac{3}{4}+\frac{\nu}{2})_{i_0}}\right.\right.\nonumber\\
 &&\times \left. \sum_{i_1=i_0}^{\infty }\frac{(\frac{3}{2}+\frac{\nu}{2})_{i_0}(\frac{5}{4}+\frac{\nu}{2})_{i_0}}{(\frac{3}{2}+\frac{\nu}{2})_{i_1}(\frac{5}{4}+\frac{\nu}{2})_{i_1}} \eta^{i_1} \right\}x
+ \sum_{n=2}^{\infty } \left\{ \sum_{i_0=0}^{\infty } \frac{(i_0+\frac{\nu}{2})^2 - \frac{1}{4^2}(\lambda +2q)}{(i_0+\frac{1}{2}+\frac{\nu}{2})(i_0+\frac{1}{4}+\frac{\nu}{2})} \frac{1}{(1+\frac{\nu}{2})_{i_0} (\frac{3}{4}+\frac{\nu}{2})_{i_0}}\right.\nonumber\\
&&\times \prod _{k=1}^{n-1} \left\{ \sum_{i_k=i_{k-1}}^{\infty }  \frac{(i_k+\frac{k}{2}+\frac{\nu}{2})^2 - \frac{1}{4^2}(\lambda +2q)}{(i_k+\frac{k}{2}+\frac{1}{2}+\frac{\nu}{2})(i_k +\frac{k}{2}+\frac{1}{4}+\frac{\nu}{2})} \frac{(1+\frac{k}{2}+ \frac{\nu}{2})_{i_{k-1}}(\frac{3}{4}+\frac{k}{2}+\frac{\nu}{2})_{i_{k-1}}}{(1+\frac{k}{2}+ \frac{\nu}{2})_{i_{k}}(\frac{3}{4}+\frac{k}{2}+\frac{\nu}{2})_{i_{k}}}\right\} \nonumber\\
&&\times \left.\left.\sum_{i_n= i_{n-1}}^{\infty } \frac{(1+\frac{n}{2}+ \frac{\nu}{2})_{i_{n-1}}(\frac{3}{4}+\frac{n}{2}+\frac{\nu}{2})_{i_{n-1}}}{(1+\frac{n}{2}+ \frac{\nu}{2})_{i_{n}}(\frac{3}{4}+\frac{n}{2}+\frac{\nu}{2})_{i_{n}}} \eta ^{i_n} \right\} x^n \right\} \hspace{1cm} \mathrm{where}\hspace{.5cm} \eta =\frac{1}{4}q x^2 \label{eq:7}
\end{eqnarray}
Put $c_0$= 1 as $\nu =0$ for the first independent solution of Mathieu equation and $\nu =\frac{1}{2}$ for the second one in (\ref{eq:7}).
\begin{rmk}
The power series expansion of Mathieu equation of the first kind for infinite series about $x=0$ using 3TRF is given by
\begin{eqnarray}
 y(x)&=& MF\left( q, \lambda; \eta = \frac{1}{4}qx^2, x=\mathrm{\cos}^2z \right) \nonumber\\
&=&  \sum_{i_0=0}^{\infty} \frac{1}{(1)_{i_0} (\frac{3}{4})_{i_0}} \eta^{i_0} +   \left\{ \sum_{i_0=0}^{\infty }\frac{i_0^2 - \frac{1}{4^2}(\lambda +2q)}{(i_0+\frac{1}{2})(i_0+\frac{1}{4})} \frac{1}{(1)_{i_0} (\frac{3}{4})_{i_0}}
 \sum_{i_1=i_0}^{\infty } \frac{(\frac{3}{2})_{i_0}(\frac{5}{4})_{i_0}}{(\frac{3}{2})_{i_1}(\frac{5}{4})_{i_1}} \eta^{i_1} \right\}x\nonumber\\
&&+ \sum_{n=2}^{\infty } \left\{ \sum_{i_0=0}^{\infty }  \frac{i_0^2 - \frac{1}{4^2}(\lambda +2q)}{(i_0+\frac{1}{2})(i_0+\frac{1}{4})} \frac{1}{(1)_{i_0} (\frac{3}{4})_{i_0}}\right. \nonumber\\
&&\times \prod _{k=1}^{n-1} \left\{ \sum_{i_k=i_{k-1}}^{\infty }  \frac{(i_k+\frac{k}{2})^2 - \frac{1}{4^2}(\lambda +2q)}{(i_k+\frac{k}{2}+\frac{1}{2})(i_k +\frac{k}{2}+\frac{1}{4})} \frac{(1+\frac{k}{2})_{i_{k-1}}(\frac{3}{4}+\frac{k}{2})_{i_{k-1}}}{(1+\frac{k}{2})_{i_{k}}(\frac{3}{4}+\frac{k}{2})_{i_{k}}}\right\} \nonumber\\
&&\times \left. \sum_{i_n= i_{n-1}}^{\infty } \frac{(1+\frac{n}{2})_{i_{n-1}}(\frac{3}{4}+\frac{n}{2})_{i_{n-1}}}{(1+\frac{n}{2})_{i_{n}}(\frac{3}{4}+\frac{n}{2})_{i_{n}}} \eta ^{i_n}\right\}  x^n  \label{eq:8}
\end{eqnarray}
\end{rmk}
\begin{rmk}
The power series expansion of Mathieu equation of the second kind for infinite series about $x=0$ using 3TRF is given by
\begin{eqnarray}
 y(x)&=& MS\left( q, \lambda; \eta = \frac{1}{4}qx^2, x=\mathrm{\cos}^2z \right) \nonumber\\
&=& x^{\frac{1}{2}} \left\{\sum_{i_0=0}^{\infty} \frac{1}{(\frac{5}{4})_{i_0} (1)_{i_0}} \eta^{i_0} \right. +  \left\{ \sum_{i_0=0}^{\infty }\frac{(i_0+\frac{1}{4})^2 - \frac{1}{4^2}(\lambda +2q)}{(i_0+\frac{3}{4})(i_0+\frac{1}{2})} \frac{1}{(\frac{5}{4})_{i_0} (1)_{i_0}}
  \sum_{i_1=i_0}^{\infty } \frac{(\frac{7}{4})_{i_0}(\frac{3}{2})_{i_0}}{(\frac{7}{4})_{i_1}(\frac{3}{2})_{i_1}} \eta^{i_1} \right\}x\nonumber\\
&&+ \sum_{n=2}^{\infty } \left\{ \sum_{i_0=0}^{\infty } \frac{(i_0+\frac{1}{4})^2 - \frac{1}{4^2}(\lambda +2q)}{(i_0+\frac{3}{4})(i_0+\frac{1}{2})} \frac{1}{(\frac{5}{4})_{i_0} (1)_{i_0}}\right.\nonumber\\
&&\times \prod _{k=1}^{n-1} \left\{ \sum_{i_k=i_{k-1}}^{\infty }  \frac{(i_k+\frac{1}{4}+\frac{k}{2})^2 - \frac{1}{4^2}(\lambda +2q)}{(i_k+\frac{3}{4}+\frac{k}{2})(i_k +\frac{1}{2}+\frac{k}{2})} \frac{(\frac{5}{4}+\frac{k}{2})_{i_{k-1}}(1+\frac{k}{2})_{i_{k-1}}}{(\frac{5}{4}+\frac{k}{2})_{i_{k}}(1+\frac{k}{2})_{i_{k}}}\right\} \nonumber\\
&&\times \left.\left. \sum_{i_n= i_{n-1}}^{\infty } \frac{(\frac{5}{4}+\frac{n}{2})_{i_{n-1}}(1+\frac{n}{2})_{i_{n-1}}}{( \frac{5}{4}+\frac{n}{2})_{i_{n}}(1+\frac{n}{2})_{i_{n}}} \eta ^{i_n}\right\} x^n \right\}\label{eq:9}
\end{eqnarray}
\end{rmk}
\section{Integral formalism for infinite series}
There is a generalized hypergeometric function which is written by
\begin{eqnarray}
G_j &=& \sum_{i_j= i_{j-1}}^{\infty } \frac{(1+\frac{j}{2}+\frac{\nu}{2})_{i_{j-1}}(\frac{3}{4}+\frac{j}{2}+\frac{\nu}{2})_{i_{j-1}}}{(1+\frac{j}{2}+\frac{\nu}{2})_{i_{j}}(\frac{3}{4}+\frac{j}{2}+\frac{\nu}{2})_{i_{j}}} \eta^{i_j}\nonumber\\
&=& \eta ^{i_{j-1}} \sum_{l=0}^{\infty } \frac{1}{(i_{j-1}+\frac{3}{4}+\frac{j}{2}+\frac{\nu}{2})_l \;(i_{j-1}+1+\frac{j}{2}+\frac{\nu }{2})_l} \eta ^l \nonumber\\
&=& \eta ^{i_{j-1}} 
\sum_{l=0}^{\infty } \frac{B(i_{j-1}-\frac{1}{4}+\frac{j}{2}+\frac{\nu}{2},l+1) B(i_{j-1}+\frac{j}{2}+\frac{\nu}{2},l+1)}{(i_{j-1}-\frac{1}{4}+\frac{j}{2}+\frac{\nu}{2})^{-1}(i_{j-1}+\frac{j}{2}+ \frac{\nu }{2})^{-1}(1)_l \;l!} \eta ^l\label{eq:21}
\end{eqnarray}
By using integral form of beta function,
\begin{subequations}
\begin{equation}
B\left(i_{j-1}-\frac{1}{4}+\frac{j}{2}+\frac{\nu}{2},l+1\right)= \int_{0}^{1} dt_j\;t_j^{i_{j-1}-\frac{5}{4}+\frac{j}{2}+\frac{\nu }{2}} (1-t_j)^l\label{eq:22a}
\end{equation}
\begin{equation}
B\left(i_{j-1}+\frac{j}{2}+\frac{\nu}{2},l+1\right)= \int_{0}^{1} du_j\;u_j^{i_{j-1}-1+\frac{j}{2}+\frac{\nu }{2}} (1-u_j)^l\label{eq:22b}
\end{equation}
\end{subequations}
Substitute (\ref{eq:22a}) and (\ref{eq:22b}) into (\ref{eq:21}). And divide $(i_{j-1}-\frac{1}{4}+\frac{j}{2}+\frac{\nu}{2})(i_{j-1}+\frac{j}{2}+ \frac{\nu }{2})$ into the new (\ref{eq:21}).
\begin{eqnarray}
K_j &=& \frac{1}{(i_{j-1}-\frac{1}{4}+\frac{j}{2}+\frac{\nu}{2})(i_{j-1}+\frac{j}{2}+ \frac{\nu }{2})}\sum_{i_j= i_{j-1}}^{\infty } \frac{(1+\frac{j}{2}+\frac{\nu}{2})_{i_{j-1}}(\frac{3}{4}+\frac{j}{2}+\frac{\nu}{2})_{i_{j-1}}}{(1+\frac{j}{2}+\frac{\nu}{2})_{i_{j}}(\frac{3}{4}+\frac{j}{2}+\frac{\nu}{2})_{i_{j}}} \eta^{i_j}\nonumber\\
&=&  \int_{0}^{1} dt_j\;t_j^{-\frac{5}{4}+\frac{j}{2}+\frac{\nu }{2}} \int_{0}^{1} du_j\;u_j^{-1+\frac{j}{2}+\frac{\nu}{2}} (\eta t_j u_j)^{i_{j-1}}
\sum_{l=0}^{\infty } \frac{1}{(1)_l \;l!} [\eta (1-t_j)(1-u_j)]^l\label{eq:23}
\end{eqnarray}
The modified Bessel function is defined by
\begin{equation}
I_{\alpha }(x)= \sum_{l=0}^{\infty } \frac{1}{l!\;(l+\alpha )!} \left( \frac{x}{2}\right)^{2l+\alpha } = \frac{\left( \frac{x}{2}\right)^{\alpha }}{\Gamma(\frac{1}{2}) \Gamma(\alpha +\frac{1}{2}) } \int_{-1}^{1} dv_j\;(1-v_j^2)^{\alpha -\frac{1}{2}} e^{-xv_j}\label{eq:24}
\end{equation}
replaced $\alpha $ and $x$ by 0 and $2\sqrt{\eta (1-t_j)(1-u_j)}$ in (\ref{eq:24}).
\begin{eqnarray}
I_0\left(2\sqrt{\eta (1-t_j)(1-u_j)}\right) &=& \sum_{l=0}^{\infty } \frac{1}{l!\;(1)_l} [\eta (1-t_j)(1-u_j)]^l\nonumber\\
 &=& \frac{1}{\pi } \int_{-1}^{1} dv_j\;(1-v_j^2)^{-\frac{1}{2}} \exp\left(-2\sqrt{\eta (1-t_j)(1-u_j)}\;v_j\right)\label{eq:25}
\end{eqnarray}
Substitute (\ref{eq:25}) into (\ref{eq:23}).
\begin{eqnarray}
K_j &=& \frac{1}{(i_{j-1}-\frac{1}{4}+\frac{j}{2}+\frac{\nu}{2})(i_{j-1}+\frac{j}{2}+ \frac{\nu }{2})}\sum_{i_j= i_{j-1}}^{\infty } \frac{(1+\frac{j}{2}+\frac{\nu}{2})_{i_{j-1}}(\frac{3}{4}+\frac{j}{2}+\frac{\nu}{2})_{i_{j-1}}}{(1+\frac{j}{2}+\frac{\nu}{2})_{i_{j}}(\frac{3}{4}+\frac{j}{2}+\frac{\nu}{2})_{i_{j}}} \eta^{i_j}\nonumber\\
&=&  \int_{0}^{1} dt_j\;t_j^{-\frac{5}{4}+\frac{j}{2}+\frac{\nu }{2}} \int_{0}^{1} du_j\;u_j^{-1+\frac{j}{2}+\frac{\nu}{2}} I_0\left(2\sqrt{\eta (1-t_j)(1-u_j)}\right) (\eta t_j u_j)^{i_{j-1}}\label{eq:26}
\end{eqnarray}
Substitute (\ref{eq:26}) into (\ref{eq:7}); apply $K_1$ into the second summation of sub-power series $y_1(x)$, apply $K_2$ into the third summation and $K_1$ into the second summation of sub-power series $y_2(x)$, apply $K_3$ into the forth summation, $K_2$ into the third summation and $K_1$ into the second summation of sub-power series $y_3(x)$, etc.\footnote{$y_1(x)$ means the sub-power series in (\ref{eq:7}) contains one term of $A_n's$, $y_2(x)$ means the sub-power series in (\ref{eq:7}) contains two terms of $A_n's$, $y_3(x)$ means the sub-power series in (\ref{eq:7}) contains three terms of $A_n's$, etc.}
\begin{thm}
The general expression of the integral representation of Mathieu equation for infinite series is given by
\begin{eqnarray}
 y(x)  &=& \sum_{n=0}^{\infty } y_{n}(x) = y_0(x)+ y_1(x)+ y_2(x)+ y_3(x)+\cdots \nonumber\\
&=& c_0 x^{\nu } \left\{ \sum_{i_0=0}^{\infty }\frac{1}{(1+\frac{\nu}{2})_{i_0}(\frac{3}{4}+\frac{\nu }{2})_{i_0}}  \eta^{i_0}\right.\nonumber\\
&& + \sum_{n=1}^{\infty } \left\{\prod _{k=0}^{n-1} \left\{ \int_{0}^{1} dt_{n-k}\;t_{n-k}^{-\frac{5}{4}+\frac{1}{2}(n-k+\nu)} \int_{0}^{1} du_{n-k}\;u_{n-k}^{-1+\frac{1}{2}(n-k+\nu)} \right.\right.\nonumber\\
&&\times I_0\left(2\sqrt{w_{n+1-k,n}(1-t_{n-k})(1-u_{n-k})}\right)\nonumber\\
&&\times  \left.\left( w_{n-k,n}^{-\frac{1}{2}(n-1-k+\nu)}\left(w_{n-k,n}\partial_{w_{n-k,n}}\right)^2 w_{n-k,n}^{\frac{1}{2}(n-1-k+\nu)}- \frac{1}{4^2}(\lambda +2q) \right) \right\}\nonumber\\
&&\times \left.\left. \sum_{i_0=0}^{\infty }\frac{1}{(1+\frac{\nu}{2})_{i_0}(\frac{3}{4}+\frac{\nu}{2})_{i_0}}  w_{1,n}^{i_0}\right\}  x^n \right\} \label{eq:28}
\end{eqnarray}
where
\begin{equation}w_{a,b}=
\begin{cases} \displaystyle {\eta \prod _{l=a}^{b} t_l u_l }\;\;\mbox{where}\; a\leq b\cr
\eta  \;\;\mbox{only}\;\mbox{if}\; a>b
\end{cases}
\label{eq:27}
\end{equation}
\end{thm}
\begin{pot} 
In (\ref{eq:7}) sub-power series $y_0(x) $, $y_1(x)$, $y_2(x)$ and $y_3(x)$ of Mathieu equation for infinite series are given by
\begin{equation}
 y(x)= \sum_{n=0}^{\infty } y_{n}(x) = y_0(x)+ y_1(x)+ y_2(x)+y_3(x)+\cdots \label{eq:100}
\end{equation}
where
\begin{subequations}
\begin{equation}
 y_0(x)= c_0 x^{\nu } \sum_{i_0=0}^{\infty} \frac{1}{(1+\frac{\nu}{2})_{i_0} (\frac{3}{4}+\frac{\nu}{2})_{i_0}} \eta^{i_0} \label{eq:101a}
\end{equation}
\begin{eqnarray}
 y_1(x)&=& c_0 x^{\nu } \Bigg\{ \sum_{i_0=0}^{\infty } \frac{(i_0+\frac{\nu}{2})^2 - \frac{1}{4^2}(\lambda +2q)}{(i_0+\frac{1}{2}+\frac{\nu}{2})(i_0+\frac{1}{4}+\frac{\nu}{2})} \frac{1}{(1+\frac{\nu}{2})_{i_0} (\frac{3}{4}+\frac{\nu}{2})_{i_0}}\nonumber\\
 &&\times \sum_{i_1=i_0}^{\infty }\frac{(\frac{3}{2}+\frac{\nu}{2})_{i_0}(\frac{5}{4}+\frac{\nu}{2})_{i_0}}{(\frac{3}{2}+\frac{\nu}{2})_{i_1}(\frac{5}{4}+\frac{\nu}{2})_{i_1}} \eta^{i_1} \Bigg\}x \label{eq:101b}
\end{eqnarray}
 \begin{eqnarray}
 y_2(x)&=& c_0 x^{\nu} \Bigg\{ \sum_{i_0=0}^{\infty }\frac{(i_0+\frac{\nu}{2})^2 - \frac{1}{4^2}(\lambda +2q)}{(i_0+\frac{1}{2}+\frac{\nu}{2})(i_0+\frac{1}{4}+\frac{\nu}{2})} \frac{1}{(1+\frac{\nu}{2})_{i_0} (\frac{3}{4}+\frac{\nu}{2})_{i_0}}\nonumber\\
&&\times  \sum_{i_1=i_0}^{\infty }  \frac{(i_1+\frac{1}{2}+\frac{\nu}{2})^2 - \frac{1}{4^2}(\lambda +2q)}{(i_1+1+\frac{\nu}{2})(i_1+\frac{3}{4}+\frac{\nu}{2})} \frac{(\frac{3}{2}+\frac{\nu}{2})_{i_0}(\frac{5}{4}+\frac{\nu}{2})_{i_0}}{(\frac{3}{2}+\frac{\nu}{2})_{i_1}(\frac{5}{4}+\frac{\nu}{2})_{i_1}} \nonumber\\
&&\times \sum_{i_2=i_1}^{\infty }  \frac{(2+\frac{\nu}{2})_{i_1}(\frac{7}{4}+\frac{\nu}{2})_{i_1}}{(2+\frac{\nu}{2})_{i_2}(\frac{7}{4}+\frac{\nu}{2})_{i_2}} \eta ^{i_2}  \Bigg\} x^2 
\label{eq:101c}
\end{eqnarray}
\begin{eqnarray}
 y_3(x)&=& c_0 x^{\nu}  \Bigg\{ \sum_{i_0=0}^{\infty }\frac{(i_0+\frac{\nu}{2})^2 - \frac{1}{4^2}(\lambda +2q)}{(i_0+\frac{1}{2}+\frac{\nu}{2})(i_0+\frac{1}{4}+\frac{\nu}{2})} \frac{1}{(1+\frac{\nu}{2})_{i_0} (\frac{3}{4}+\frac{\nu}{2})_{i_0}}\nonumber\\
&&\times  \sum_{i_1=i_0}^{\infty }\frac{(i_1+\frac{1}{2}+\frac{\nu}{2})^2 - \frac{1}{4^2}(\lambda +2q)}{(i_1+1+\frac{\nu}{2})(i_1+\frac{3}{4}+\frac{\nu}{2})} \frac{(\frac{3}{2}+\frac{\nu}{2})_{i_0}(\frac{5}{4}+\frac{\nu}{2})_{i_0}}{(\frac{3}{2}+\frac{\nu}{2})_{i_1}(\frac{5}{4}+\frac{\nu}{2})_{i_1}}  \nonumber\\
&&\times \sum_{i_2=i_1}^{\infty }\frac{(i_2+1+\frac{\nu}{2})^2 - \frac{1}{4^2}(\lambda +2q)}{(i_2+\frac{3}{2}+\frac{\nu}{2})(i_2+\frac{5}{4}+\frac{\nu}{2})} \frac{(2+\frac{\nu}{2})_{i_1}(\frac{7}{4}+\frac{\nu}{2})_{i_1}}{(2+\frac{\nu}{2})_{i_2}(\frac{7}{4}+\frac{\nu}{2})_{i_2}} \nonumber\\
 &&\times \sum_{i_3=i_2}^{\infty }\frac{(\frac{5}{2}+\frac{\nu}{2})_{i_2}(\frac{9}{4}+\frac{\nu}{2})_{i_2}}{(\frac{5}{2}+\frac{\nu}{2})_{i_3}(\frac{9}{4}+\frac{\nu}{2})_{i_3}} \eta ^{i_3} \Bigg\}x^3
\label{eq:101d}
\end{eqnarray}
\end{subequations}
Put $j=1$ in (\ref{eq:26}). Take the new (\ref{eq:26}) into (\ref{eq:101b}).
\begin{eqnarray}
 y_1(x)&=& c_0 x^{\nu } \int_{0}^{1} dt_1\;t_1^{-\frac{3}{4}+\frac{\nu }{2}} \int_{0}^{1} du_1\;u_1^{-\frac{1}{2}+\frac{\nu}{2}} I_0\left(2\sqrt{\eta (1-t_1)(1-u_1)}\right) \nonumber\\
&&\times  \Bigg\{ \sum_{i_0=0}^{\infty } \left( \left(i_0+\frac{\nu}{2}\right)^2 -\frac{1}{4^2}(\lambda +2q)\right) \frac{1}{(1+\frac{\nu}{2})_{i_0}(\frac{3}{4}+\frac{\nu}{2})_{i_0}} (\eta t_1 u_1)^{i_0} \Bigg\}x\nonumber\\
&=& c_0 x^{\nu } \int_{0}^{1} dt_1\;t_1^{-\frac{3}{4}+\frac{\nu }{2}} \int_{0}^{1} du_1\;u_1^{-\frac{1}{2}+\frac{\nu}{2}} I_0\left(2\sqrt{\eta (1-t_1)(1-u_1)}\right) \nonumber\\
&&\times  \left( w_{1,1}^{-\frac{\nu}{2}}\left(w_{1,1}\partial_{w_{1,1}}\right)^2 w_{1,1}^{\frac{\nu}{2}}- \frac{1}{4^2}(\lambda +2q) \right) \Bigg\{ \sum_{i_0=0}^{\infty } \frac{1}{(1+\frac{\nu}{2})_{i_0}(\frac{3}{4}+\frac{\nu}{2})_{i_0}} w_{1,1}^{i_0}\Bigg\}x \label{eq:102}\\
&& \mathrm{where}\hspace{.5cm} w_{1,1}=\eta \prod _{l=1}^{1} t_l u_l \nonumber
\end{eqnarray}
Put $j=2$ in (\ref{eq:26}). Take the new (\ref{eq:26}) into (\ref{eq:101c}). 
\begin{eqnarray}
 y_2(x)&=& c_0 x^{ \nu } \int_{0}^{1} dt_2\;t_2^{-\frac{1}{4}+\frac{\nu }{2}} \int_{0}^{1} du_2\;u_2^{\frac{\nu}{2}} I_0\left(2\sqrt{\eta (1-t_2)(1-u_2)}\right)\nonumber\\
&&\times \left( w_{2,2}^{-(\frac{1}{2}+\frac{\nu}{2})}\left(w_{2,2}\partial_{w_{2,2}}\right)^2 w_{2,2}^{\frac{1}{2}+\frac{\nu}{2}}- \frac{1}{4^2}(\lambda +2q) \right)\nonumber\\
&&\times \Bigg\{ \sum_{i_0=0}^{\infty }\frac{(i_0+\frac{\nu}{2})^2 - \frac{1}{4^2}(\lambda +2q)}{(i_0+\frac{1}{2}+\frac{\nu}{2})(i_0+\frac{1}{4}+\frac{\nu}{2})} \frac{1}{(1+\frac{\nu}{2})_{i_0} (\frac{3}{4}+\frac{\nu}{2})_{i_0}}\nonumber\\
&&\times \sum_{i_1=i_0}^{\infty } \frac{(\frac{3}{2}+\frac{\nu}{2})_{i_0}(\frac{5}{4}+\frac{\nu}{2})_{i_0}}{(\frac{3}{2}+\frac{\nu}{2})_{i_1}(\frac{5}{4}+\frac{\nu}{2})_{i_1}} w_{2,2}^{i_1} \Bigg\}x^2  \label{eq:103}\\
&& \mathrm{where}\hspace{.5cm} w_{2,2}=\eta \prod _{l=2}^{2} t_l u_l \nonumber
\end{eqnarray}
Put $j=1$ and $\eta  = w_{2,2}$ in (\ref{eq:26}). Take the new (\ref{eq:26}) into (\ref{eq:103}).
\begin{eqnarray}
 y_2(x)&=& c_0 x^{\nu } \int_{0}^{1} dt_2\;t_2^{-\frac{1}{4}+\frac{\nu }{2}} \int_{0}^{1} du_2\;u_2^{\frac{\nu}{2}} I_0\left(2\sqrt{\eta (1-t_2)(1-u_2)}\right) \nonumber\\
&&\times \left( w_{2,2}^{-(\frac{1}{2}+\frac{\nu}{2})}\left(w_{2,2}\partial_{w_{2,2}}\right)^2 w_{2,2}^{\frac{1}{2}+\frac{\nu}{2}}- \frac{1}{4^2}(\lambda +2q) \right)\nonumber\\
&&\times  \int_{0}^{1} dt_1\;t_1^{-\frac{3}{4}+\frac{\nu }{2}} \int_{0}^{1} du_1\;u_1^{-\frac{1}{2}+\frac{\nu}{2}} I_0\left(2\sqrt{w_{2,2}(1-t_1)(1-u_1)}\right)\nonumber\\
&&\times \left( w_{1,2}^{-\frac{\nu}{2}}\left(w_{1,2}\partial_{w_{1,2}}\right)^2 w_{1,2}^{\frac{\nu}{2}}- \frac{1}{4^2}(\lambda +2q) \right)\nonumber\\
&&\times  \Bigg\{ \sum_{i_0=0}^{\infty } \frac{1}{(1+\frac{\nu}{2})_{i_0}(\frac{3}{4}+\frac{\nu}{2})_{i_0}} w_{1,2}^{i_0} \Bigg\}x^2 \label{eq:104}\\
&& \mathrm{where}\hspace{.5cm} w_{1,2}=\eta \prod _{l=1}^{2} t_l u_l \nonumber 
\end{eqnarray}
By using similar process for the previous cases of integral forms of $y_1(x)$ and $y_2(x)$, the integral form of sub-power series expansion of $y_3(x)$ is
\begin{eqnarray}
 y_3(x)&=& c_0 x^{\nu } \int_{0}^{1} dt_3\;t_3^{\frac{1}{4}+\frac{\nu }{2}} \int_{0}^{1} du_3\;u_3^{\frac{1}{2}+\frac{\nu}{2}} I_0\left(2\sqrt{\eta (1-t_3)(1-u_3)}\right) \nonumber\\
&&\times\left( w_{3,3}^{-(1+\frac{\nu}{2})}\left(w_{3,3}\partial_{w_{3,3}}\right)^2 w_{3,3}^{1+\frac{\nu}{2}}- \frac{1}{4^2}(\lambda +2q) \right) \nonumber\\
&&\times \int_{0}^{1} dt_2\;t_2^{-\frac{1}{4}+\frac{\nu }{2}} \int_{0}^{1} du_2\;u_2^{\frac{\nu}{2}} I_0\left(2\sqrt{w_{3,3}(1-t_2)(1-u_2)}\right) \nonumber\\
&&\times\left( w_{2,3}^{-(\frac{1}{2}+\frac{\nu}{2})}\left(w_{2,3}\partial_{w_{2,3}}\right)^2 w_{2,3}^{\frac{1}{2}+\frac{\nu}{2}}- \frac{1}{4^2}(\lambda +2q) \right) \nonumber\\
&&\times  \int_{0}^{1} dt_1\;t_1^{-\frac{3}{4}+\frac{\nu }{2}} \int_{0}^{1} du_1\;u_1^{-\frac{1}{2}+\frac{\nu}{2}} I_0\left(2\sqrt{w_{2,3}(1-t_1)(1-u_1)}\right)\nonumber\\
&&\times \left( w_{1,3}^{-\frac{\nu}{2}}\left(w_{1,3}\partial_{w_{1,3}}\right)^2 w_{1,3}^{\frac{\nu}{2}}- \frac{1}{4^2}(\lambda +2q) \right) \nonumber\\
&&\times \Bigg\{\sum_{i_0=0}^{\infty } \frac{1}{(1+\frac{\nu}{2})_{i_0} (\frac{3}{4}+\frac{\nu}{2})_{i_0}}w_{1,3}^{i_0} \Bigg\} x^3 \label{eq:105}
\end{eqnarray}
where
\begin{equation}
w_{3,3}=\eta \prod _{l=3}^{3} t_l u_l \hspace{1cm}
w_{2,3}=\eta \prod _{l=2}^{3} t_l u_l \hspace{1cm}
w_{1,3}= \eta \prod _{l=1}^{3} t_l u_l
\nonumber
\end{equation}
By repeating this process for all higher terms of integral forms of sub-summation $y_m(x)$ terms where $m \geq 4$, we obtain every integral forms of $y_m(x)$ terms. 
Since we substitute (\ref{eq:101a}), (\ref{eq:102}), (\ref{eq:104}), (\ref{eq:105}) and including all integral forms of $y_m(x)$ terms where $m \geq 4$ into (\ref{eq:100}), we obtain (\ref{eq:28}).
\qed
\end{pot}
Put $c_0= 1$ and $\nu = 0$ in (\ref{eq:28}).
\begin{eqnarray}
 y(x)&=& MF\left( q, \lambda; x=\mathrm{\cos}^2z, \eta = \frac{1}{4}qx^2 \right) \nonumber\\
 &=&  \sum_{i_0=0}^{\infty }\frac{1}{(1)_{i_0}(\frac{3}{4})_{i_0}}  \eta^{i_0} 
 + \sum_{n=1}^{\infty } \left\{\prod _{k=0}^{n-1} \left\{ \int_{0}^{1} dt_{n-k}\;t_{n-k}^{-\frac{5}{4}+\frac{1}{2}(n-k)} \int_{0}^{1} du_{n-k}\;u_{n-k}^{-1+\frac{1}{2}(n-k)} \right.\right.\nonumber\\
&&\times I_0\left(2\sqrt{w_{n+1-k,n}(1-t_{n-k})(1-u_{n-k})}\right)\label{eq:29}\\
&&\times  \left.\left( w_{n-k,n}^{-\frac{1}{2}(n-1-k)}\left(w_{n-k,n}\partial_{w_{n-k,n}}\right)^2 w_{n-k,n}^{\frac{1}{2}(n-1-k)}- \frac{1}{4^2}(\lambda +2q) \right)\right\}
 \left. \sum_{i_0=0}^{\infty }\frac{1}{(1)_{i_0}(\frac{3}{4})_{i_0}}  w_{1,n}^{i_0}\right\} x^n  \nonumber
\end{eqnarray}
replaced $\alpha $ and $x$ by $-\frac{1}{4}$ and $2\sqrt{\eta}$ in (\ref{eq:24}).
\begin{equation}
I_{-\frac{1}{4}}\left(2\sqrt{\eta}\right) = \frac{\eta ^{-\frac{1}{8}}}{\Gamma (\frac{3}{4})}\sum_{l=0}^{\infty } \frac{1}{l!\;(\frac{3}{4})_l} \eta ^l =\frac{\eta ^{-\frac{1}{8}}}{\Gamma (\frac{1}{2})\Gamma (\frac{1}{4})} \int_{-1}^{1} dv_0\;(1-v_0^2)^{-\frac{3}{4}} \exp\left(-2\sqrt{\eta}\;v_0\right)\label{eq:30}
\end{equation}
Similarly,
\begin{eqnarray}
I_{-\frac{1}{4}}\left(2\sqrt{w_{1,n}}\right) &=& \frac{w_{1,n}^{-\frac{1}{8}}}{\Gamma (\frac{3}{4})}\sum_{l=0}^{\infty } \frac{1}{l!\;(\frac{3}{4})_l} w_{1,n} ^l =\frac{w_{1,n}^{-\frac{1}{8}}}{\Gamma (\frac{1}{2})\Gamma (\frac{1}{4})} \int_{-1}^{1} dv_0\;(1-v_0^2)^{-\frac{3}{4}} \exp\left(-2\sqrt{w_{1,n}}\;v_0\right)\hspace{1cm}\label{eq:31}
\end{eqnarray}
Substitute (\ref{eq:30}) and (\ref{eq:31}) into (\ref{eq:29}).
\begin{rmk}
The integral representation of Mathieu equation of the first kind for infinite series about $x=0$ using 3TRF is given by
\begin{eqnarray}
 y(x)&=& MF\left( q, \lambda; x=\mathrm{\cos}^2z, \eta = \frac{1}{4}qx^2 \right) \nonumber\\
 &=& \Gamma \left( 3/4 \right) \Bigg\{ \eta ^{\frac{1}{8}} I_{-\frac{1}{4}}\left(2\sqrt{\eta}\right)+ \sum_{n=1}^{\infty } \Bigg\{\prod _{k=0}^{n-1} \left\{ \int_{0}^{1} dt_{n-k}\;t_{n-k}^{-\frac{5}{4}+\frac{1}{2}(n-k)} \int_{0}^{1} du_{n-k}\;u_{n-k}^{-1+\frac{1}{2}(n-k)}\right. \nonumber\\
&&\times  I_0\left(2\sqrt{w_{n+1-k,n}(1-t_{n-k})(1-u_{n-k})}\right)\nonumber\\
&&\times  \left. \left( w_{n-k,n}^{-\frac{1}{2}(n-1-k)}\left(w_{n-k,n}\partial_{w_{n-k,n}}\right)^2 w_{n-k,n}^{\frac{1}{2}(n-1-k)}- \frac{1}{4^2}(\lambda +2q) \right)\right\}
  w_{1,n}^{\frac{1}{8}} I_{-\frac{1}{4}}\left(2\sqrt{w_{1,n}}\right) \Bigg\} x^n \Bigg\} \label{eq:32}
\end{eqnarray}
\end{rmk}
Put $c_0= 1$ and $\nu = \frac{1}{2}$ in (\ref{eq:28}).
\begin{eqnarray}
 y(x)&=& MS\left( q, \lambda; x=\mathrm{\cos}^2z, \eta = \frac{1}{4}qx^2 \right) \nonumber\\
&=& x^{\frac{1}{2}} \Bigg\{ \sum_{i_0=0}^{\infty }\frac{1}{(\frac{5}{4})_{i_0}(1)_{i_0}}  \eta^{i_0}
+ \sum_{n=1}^{\infty } \Bigg\{\prod _{k=0}^{n-1} \left\{ \int_{0}^{1} dt_{n-k}\;t_{n-k}^{-1+\frac{1}{2}(n-k)} \int_{0}^{1} du_{n-k}\;u_{n-k}^{-\frac{3}{4}+\frac{1}{2}(n-k)} \right.\nonumber\\
&&\times I_0\left(2\sqrt{w_{n+1-k,n}(1-t_{n-k})(1-u_{n-k})}\right)\label{eq:33}\\
&&\times \left. \left( w_{n-k,n}^{-\frac{1}{2}(n-k-\frac{1}{2})}\left(w_{n-k,n}\partial_{w_{n-k,n}}\right)^2 w_{n-k,n}^{\frac{1}{2}(n-k-\frac{1}{2})}- \frac{1}{4^2}(\lambda +2q) \right)\right\} 
 \sum_{i_0=0}^{\infty }\frac{1}{(\frac{5}{4})_{i_0}(1)_{i_0}}  w_{1,n}^{i_0}\Bigg\}  x^n \Bigg\} \nonumber
\end{eqnarray}
replaced $\alpha $ and $x$ by $\frac{1}{4}$ and $2\sqrt{\eta}$ in (\ref{eq:24}).
\begin{equation}
I_{\frac{1}{4}}\left(2\sqrt{\eta}\right) = \frac{\eta ^{\frac{1}{8}}}{\Gamma (\frac{5}{4})}\sum_{l=0}^{\infty } \frac{1}{l!\;(\frac{5}{4})_l} \eta ^l =\frac{\eta ^{\frac{1}{8}}}{\Gamma (\frac{1}{2})\Gamma (\frac{3}{4})} \int_{-1}^{1} dv_0\;(1-v_0^2)^{-\frac{1}{4}} \exp\left(-2\sqrt{\eta}\;v_0\right)\label{eq:34}
\end{equation}
Similarly,
\begin{equation}
I_{\frac{1}{4}}\left(2\sqrt{w_{1,n}}\right) = \frac{w_{1,n} ^{\frac{1}{8}}}{\Gamma (\frac{5}{4})}\sum_{l=0}^{\infty } \frac{1}{l!\;(\frac{5}{4})_l} w_{1,n} ^l = \frac{w_{1,n}^{\frac{1}{8}}}{\Gamma (\frac{1}{2})\Gamma (\frac{3}{4})} \int_{-1}^{1} dv_0\;(1-v_0^2)^{-\frac{1}{4}} \exp\left(-2\sqrt{w_{1,n}}\;v_0\right) \label{eq:35}
\end{equation}
Substitute (\ref{eq:34}) and (\ref{eq:35}) into (\ref{eq:33}).
\begin{rmk}
The integral representation of Mathieu equation of the second kind for infinite series about $x=0$ using 3TRF is given by
\begin{eqnarray}
 y(x)&=& MS\left( q, \lambda; x=\mathrm{\cos}^2z, \eta = \frac{1}{4}qx^2 \right) \nonumber\\
&=& \Gamma ( 5/4 ) x^{\frac{1}{2}} \Bigg\{ \eta ^{-\frac{1}{8}}I_{\frac{1}{4}}\left(2\sqrt{\eta}\right)+ \sum_{n=1}^{\infty } \Bigg\{\prod _{k=0}^{n-1} \left\{ \int_{0}^{1} dt_{n-k}\;t_{n-k}^{-1+\frac{1}{2}(n-k)} \int_{0}^{1} du_{n-k}\;u_{n-k}^{-\frac{3}{4}+\frac{1}{2}(n-k)} \right.\nonumber\\
&&\times I_0\left(2\sqrt{w_{n+1-k,n}(1-t_{n-k})(1-u_{n-k})}\right)\label{eq:36}\\
&&\times \left. \left( w_{n-k,n}^{-\frac{1}{2}(n-k-\frac{1}{2})}\left(w_{n-k,n}\partial_{w_{n-k,n}}\right)^2 w_{n-k,n}^{\frac{1}{2}(n-k-\frac{1}{2})}- \frac{1}{4^2}(\lambda +2q) \right)\right\} 
  w_{1,n}^{-\frac{1}{8}} I_{\frac{1}{4}}\left(2\sqrt{w_{1,n}}\right) \Bigg\}  x^n \Bigg\} \nonumber
\end{eqnarray}
\end{rmk}
 As we see (\ref{eq:32}) and (\ref{eq:36}), modified Bessel function recurs in each of sub-integral forms. We can transform the   Mathieu function from these integral forms to other well-known special functions: Kummer function, Legendre function, Hypergeometric function, Laguerre function etc. 

\section{Asymptotic behavior of the function $y(x)$ and the boundary condition for $x= \cos^2z$ for infinite series}
As $n\gg 1$ (for sufficiently large), (\ref{eq:5a}) and (\ref{eq:5b}) are
\begin{subequations}
\begin{equation}
\lim_{n\gg 1} A_n = A=1
\label{psy:1a}
\end{equation}
And,
\begin{equation}
\lim_{n\gg 1} B_n = B=\frac{q}{n^2}
\label{psy:1b}
\end{equation}
\end{subequations}
As $n\gg 1$, (\ref{psy:1b}) is extremely smaller than (\ref{psy:1a}). Put (\ref{psy:1a}) with $B_n =0$ into (\ref{eq:4}). 
\begin{equation}
c_{n+1} = c_n
\label{psy:6}
\end{equation}
Plug (\ref{psy:6}) into the power series expansion where $ \sum_{n=0}^{\infty } c_n x^n $, putting $c_0= 1 $ for simplicity.
\begin{equation}
\lim_{n\gg 1}y(x) = \frac{1}{1-x} \hspace{1cm}\mbox{where}\;0\leq x< 1
\label{psy:7}
\end{equation}
For being convergent of $y(x)$ in (\ref{psy:7}), an independent variable $x=\cos^2z$ should be less than 1. 
In general, in various areas of the modern physics, we require that a function $y(x)$ should be convergent at $x=1$.
Mathieu polynomial which make $A_n$ term terminated merely makes possible that a $y(x)$ is convergent at $x=1$, and its formal series solutions are available in chapter 7 of Ref.\cite{Choun2013}.
 
I show the power series expansion in closed forms of Mathieu equation for infinite series in this paper analytically. Also, I derive integral forms of the Mathieu function from its power series expansion. It is quiet important that a modified Bessel function recurs in each of sub-integral forms of the Mathieu function, because we can investigate how this function is associated with other well known special functions such as Bessel, Kummer, hypergeometric and Laguerre functions, etc. In future papers I will derive the Mathieu function for polynomial which makes $A_n$ term terminated by using similar methods what I do in this paper: (1) power series expansion in closed forms, (2) its integral representation and (3) its generating function.  And I will construct orthogonal relations of the Mathieu polynomial, normalized physical factors and expectation values of any physical quantities from the generating function for the Mathieu polynomial analytically. 
\section{Application}
1. By using the methods on the above, we can apply the power series expansion of Mathieu equation and its integral forms into various modern physics areas.\cite{McLa1947,Guti2003,Daym1955,Troe1973,Alha1995,Shen1981,Bhat1988,Ragh1991}
 For example, in general Mathieu equation arises from two-dimensional vibrational problems in elliptical coordinates with physical points of a view\cite{Wang1989}. Its equation is derived from the Helmholtz equation in elliptic cylinder coordinates by using the method of separation of variables (see (5)-(7) in p.610 in Ref.\cite{Wang1989}). Using the power series expansion of Mathieu equation, it might be possible to obtain specific eigenvalues for the wave equation in vibrational systems. 
The normalized constant for wave functions and its expectation values for the entire region might be also possible to be constructed by applying the generating function for the Mathieu polynomial which makes $A_n$ term terminated.\footnote{several authors treat the solution of Mathieu equation as an polynomial. In future papers I will construct the Mathieu function for polynomial which makes $A_n$ term terminated. Indeed, the generating function of it will be derived in mathematical rigour. In this paper I show analytic solutions of Mathieu equation for infinite series}
\vspace{5mm}

2. In ``Examples of Heun and Mathieu functions as solutions of wave equations in curved spaces''\cite{Birk2007}, ``Dirac equation in the background of the Nutku helicoid metric''\cite{Hort2007}, two authors consider the Dirac equation in the background of the Nutku helicoid metric in five dimensions. They obtain solutions for the four different components by using the Newman-Penrose formalism\cite{Newm1962}(see (8a)-(8d) Ref.\cite{Hort2007}). And they separate uncoupled equations for the lower components into two ordinary differential equations: angular equation is of the Mathieu type and radial equation is of the double confluent form which can be reduced to the Mathieu equation (see (11), (13) Ref.\cite{Hort2007}). 
Using power series expansion and its integral representation of Mathieu equation, it might be possible to obtain eigenvalues and normalized wave functions at various regions.\footnote{In the future I will construct analytic solutions of the Mathieu polynomial and its eigenvalues of these two problems.}  
\section{Series ``Special functions and three term recurrence formula (3TRF)''} 

This paper is 5th out of 10.
\vspace{3mm}

1. ``Approximative solution of the spin free Hamiltonian involving only scalar potential for the $q-\bar{q}$ system'' \cite{chou2012a} - In order to solve the spin-free Hamiltonian with light quark masses we are led to develop a totally new kind of special function theory in mathematics that generalize all existing theories of confluent hypergeometric types. We call it the Grand Confluent Hypergeometric Function. Our new solution produces previously unknown extra hidden quantum numbers relevant for description of supersymmetry and for generating new mass formulas.
\vspace{3mm}

2. ``Generalization of the three-term recurrence formula and its applications'' \cite{chou2012b} - Generalize three term recurrence formula in linear differential equation.  Obtain the exact solution of the three term recurrence for polynomials and infinite series.
\vspace{3mm}

3. ``The analytic solution for the power series expansion of Heun function'' \cite{chou2012c} -  Apply three term recurrence formula to the power series expansion in closed forms of Heun function (infinite series and polynomials) including all higher terms of $A_n$'s.
\vspace{3mm}

4. ``Asymptotic behavior of Heun function and its integral formalism'', \cite{Chou2012d} - Apply three term recurrence formula, derive the integral formalism, and analyze the asymptotic behavior of Heun function (including all higher terms of $A_n$'s). 
\vspace{3mm}

5. ``The power series expansion of Mathieu function and its integral formalism'', \cite{Chou2012e} - Apply three term recurrence formula, analyze the power series expansion of Mathieu function and its integral forms.  
\vspace{3mm}

6. ``Lame equation in the algebraic form'' \cite{Chou2012f} - Applying three term recurrence formula, analyze the power series expansion of Lame function in the algebraic form and its integral forms.
\vspace{3mm}

7. ``Power series and integral forms of Lame equation in   Weierstrass's form and its asymptotic behaviors'' \cite{Chou2012g} - Applying three term recurrence formula, derive the power series expansion of Lame function in   Weierstrass's form and its integral forms. 
\vspace{3mm}

8. ``The generating functions of Lame equation in   Weierstrass's form'' \cite{Chou2012h} - Derive the generating functions of Lame function in   Weierstrass's form (including all higher terms of $A_n$'s).  Apply integral forms of Lame functions in   Weierstrass's form.
\vspace{3mm}

9. ``Analytic solution for grand confluent hypergeometric function'' \cite{Chou2012i} - Apply three term recurrence formula, and formulate the exact analytic solution of grand confluent hypergeometric function (including all higher terms of $A_n$'s). Replacing $\mu $ and $\varepsilon \omega $ by 1 and $-q$, transforms the grand confluent hypergeometric function into Biconfluent Heun function.
\vspace{3mm}

10. ``The integral formalism and the generating function of grand confluent hypergeometric function'' \cite{Chou2012j} - Apply three term recurrence formula, and construct an integral formalism and a generating function of grand confluent hypergeometric function (including all higher terms of $A_n$'s). 
\section*{Acknowledgment}
I thank Bogdan Nicolescu. The discussions I had with him on number theory was of great joy.

\end{document}